\documentclass[useAMS,usenatbib]{mn2e}
\usepackage{graphicx,amssym}
\newcommand{\bc}{\begin{center}}
\newcommand{\ec}{\end{center}}
\title[Post-starburst galaxies and the build-up of the red-sequence]
      {On the role of the post-starburst phase in the build-up of the
        red-sequence of intermediate redshift clusters}
\author[G.~De Lucia et al.]
       {Gabriella De Lucia$^{1,2}$\thanks{Email: delucia@oats.inaf.it}, Bianca
         M. Poggianti$^3$, Claire Halliday$^4$, Bo Milvang-Jensen$^{5,6}$, 
         \newauthor Stefan Noll$^7$, Ian Smail$^8$, Dennis Zaritsky$^9$  
        \\  
        $^1$INAF - Astronomical Observatory of Trieste, via G.B. Tiepolo 11, 
        I-34143 Trieste, Italy\\
        $^2$Max--Planck--Institut f\"ur Astrophysik, 
        Karl--Schwarzschild--Str. 1, D-85748 Garching, Germany\\
        $^3$INAF - Astronomical Observatory of Padova, Vicolo dell'Osservatorio
        5, I-35122 Padova, Italy\\
        $^4$INAF, Astronomical Observatory of Arcetri, Largo Enrico Fermi 5,
        I-50125 Firenze, Italy\\
        $^5$Dark Cosmology Centre, Niels Bohr Institute, University of
        Copenhagen, Juliane Maries Vej 30, DK-2100 Copenhagen, Denmark\\ 
        $^6$The Royal Library / Copenhagen University Library, Research
        Department, Box 2149, DK-1016 Copenhagen K, Denmark\\
        $^7$Observatoire Astronomique de Marseille-Provence, 38 rue Fr\'ed\'eric
        Joliot-Curie, 13388 Marseille Cedex 13, France \\
        $^8$Institute for Computational Cosmology, Durham University, South
        Road, Durham DH1 3LE\\
        $^9$Steward Observatory, University of Arizona, 933 North Cherry
        Avenue, Tucson, AZ-85721, USA} 
\begin{document}

\pagerange{\pageref{firstpage}--\pageref{lastpage}} 
\pubyear{2008}

\maketitle

\label{firstpage}

\begin{abstract}
  We present new deep spectroscopic observations of 0.05--0.5 L$_*$ galaxies in
  one cluster (cl$1232.5$--$1250$) drawn from the ESO Distant Cluster Survey
  (EDisCS) sample, at $z = 0.54$. The new data extend the spectroscopy already
  available for this cluster by about 1 magnitude. The cluster has a large
  fraction of passive galaxies and exhibits a well defined and relatively tight
  colour-magnitude relation. Among spectroscopic members, only six galaxies are
  classified as `post-starburst' (k+a). For another EDisCS cluster at similar
  redshift and with as deep spectroscopy, no member is found to have a k+a
  spectrum. The low measured numbers of post-starburst systems appear to be
  inadequate to explain the observed increase of faint red galaxies at lower
  redshift, even when accounting for the infall of new galaxies onto the
  cluster.  Post-starburst galaxies represent a possible channel to move
  galaxies from the blue star-forming cloud to the red-sequence, but the
  available data suggest this is not the dominant channel in galaxy
  clusters. If the galaxies at the faint end of the red-sequence in nearby
  clusters originate from the blue galaxies observed in distant galaxy
  clusters, the transformation must have occurred primarily through physical
  processes that do not lead to a post-starburst phase. In addition, our data
  exclude a large population of k+a galaxies at faint magnitudes, as found in
  the Coma cluster.
\end{abstract}


\section{Introduction}
\label{sec:intro}

Early-type galaxies in clusters show a clearly defined relation between colour
and luminosity. The existence of this relation has been known for a long time
\citep{Baum_1959,deVaucouleurs_1961,Visvanathan_Sandage_1977} and it has long
been thought to encode important information about the formation history of
early-type galaxies. The scatter around the colour-magnitude relation is very small,
at all redshifts where it has been measured accurately, and its slope does not
appear to have evolved significantly up to $z \sim 1$ \citep[][and references
  therein]{Ellis_etal_1997,Mei_etal_2008}.

The traditional interpretation of these observations is that cluster early-type
galaxies formed at high redshift ($z > 2-3$) in a short duration event, and
evolved passively thereafter. In this scenario, the colour-magnitude relation
represents a sequence of increasing metallicity with increasing luminosity, and
is produced because a galactic wind is induced progressively later in brighter
galaxies, which have deeper potential wells
\citep{Faber_1973,Arimoto_Yoshii_1987,Kodama_Arimoto_1997}. Weak age trends are
not excluded and are indeed measured
(e.g. \citealt*{Ferreras_Charlot_Silk_1999}; \citealt{Poggianti_etal_2001};
\citealt{Nelan_etal_2005}), but age is not traditionally considered the main
driver of the colour-magnitude relation because younger stellar populations
have more rapid colour evolution. If the colour-magnitude relation was indeed
an age sequence, it would change dramatically with increasing redshift as
smaller galaxies approach their formation epochs \citep{Kodama_etal_1998}.

This classical interpretation relies on the assumption that all galaxies
sitting on the colour-magnitude relation today, can be identified as
red-sequence members of high-redshift galaxy clusters - an assumption that is
likely to be \emph{wrong}. As noticed by \citet{vanDokkum_Franx_1996},
evolutionary studies based on observations of early-type galaxies at different
epochs are seriously complicated by what they named the \emph{progenitor bias}:
if the progenitors of some present early-types were spirals at higher redshift,
they would not be included in the samples studied at higher redshift, biasing
the measured population towards older ages. The existence of such a bias, as
discussed below, has been directly demonstrated in recent studies of the
colour-magnitude evolution.

In \citet{DeLucia_etal_2004}, we analysed the colour-magnitude relation of 4
clusters in the redshift range 0.7-0.8 from the ESO Distant Cluster Survey
(EDisCS), and found a deficiency of low-luminosity red-sequence galaxies with
respect to the nearby cluster Coma. Similar results were found by
\citet{Kodama_etal_2004} for early-type galaxies in a single deep field, and
were later confirmed by a number of studies using independent datasets
\citep{DeLucia_etal_2007,Stott_etal_2007,Gilbank_etal_2008}. The reported
results are not unanimous and were criticised in particular by
\citet{Andreon_2008} whose results are, however, consistent with ours over the
same redshift range (see their Fig.~4). In addition, some evolution in the
distribution of red-sequence galaxies is expected as a consequence of the
established mass-dependence of early-type galaxy evolution \citep[e.g.][but see
  \citealt{Fontanot_etal_2009}]{Thomas_etal_2005,Nelan_etal_2005,Gallazzi_etal_2006}.

The population of blue galaxies observed in distant galaxy clusters provides
the logical progenitors of the increasing number of faint red galaxies
observed at low redshift. In \citet{DeLucia_etal_2007}, we showed that a
simple scenario in which infalling galaxies have their star formation histories
truncated by the hostile cluster environment is consistent with the observed
build-up of the colour-magnitude sequence (see also
\citealt{Smail_etal_1998}). Which physical mechanism(s) is(are) responsible for
the transformation from star-forming to passive galaxies, as well as the
associated time-scale(s) remain, however, to be understood.

While the integrated colours of any new addition to the colour-magnitude
relation indicate no sign of ongoing star formation, it is possible to apply
more sensitive spectroscopic tests to identify the signature of their recent
past activity. `K+a' (also referred to as `E+A' or `post-starburst') galaxies,
in particular, represent the best candidates for an evolutionary link between
star-forming, gas-rich galaxies and quiescent, gas-poor systems. These galaxies
are commonly identified on the basis of their optical spectra as having almost
no emission lines (typically [OII]$3727$) and exceptionally strong Balmer lines
in absorption
\citep{Dressler_Gunn_1983,Couch_Sharples_1987}. Spectrophotometric modelling
has demonstrated that the superposition of a young stellar population
(represented by A stars) and an old population (dominated by K stars) indicates
that star formation ceased abruptly in these galaxies in the past $5\times
10^7$--$1.5\times10^9$ years (see e.g. \citealt{Couch_Sharples_1987};
\citealt*{Newberry_Boroson_Kirshner_1990}; \citealt{Poggianti_Barbaro_1997};
\citealt*{Bekki_Shioya_Couch_2001}). First studied in galaxy clusters, k+a
galaxies have also been found in low-density regions
\citep{Zabludoff_etal_1996,Blake_etal_2004,Yan_etal_2008,Wild_etal_2008}
suggesting that interactions with the `cluster environment' (e.g. ram-pressure
stripping or harassment) are a \emph{possible} but not a \emph{necessary}
condition for galaxies to pass through this evolutionary phase. The
\emph{absolute} fraction of k+a galaxies in different environments and at
various cosmic epochs is still a matter of debate, depending significantly on
the criteria for Balmer and [OII] strengths, the line measurement method, the
spectral quality, and the `environment' definition adopted \citep[][and
  references therein]{Poggianti_etal_2009}. At low redshift, k+a galaxies are
rare among luminous galaxies, in all environments. For the Coma cluster, a
conspicuous population of k+a galaxies has been detected at fainter magnitudes,
representing a significant fraction of the cluster `dwarfs'
\citep{Poggianti_etal_2004}. Spectroscopic studies of distant cluster galaxy
populations are unavoidably limited to the bright end of the galaxy luminosity
function. Therefore, the presence of a large fraction of k+a galaxies at faint
magnitudes in intermediate redshift galaxy clusters, is yet to be demonstrated.

To measure how much of the build-up of the colour-magnitude relation at
low-redshift is caused by galaxies passing through a k+a phase, we have
undertaken deep spectroscopic observations of 0.05--0.5 L$_*$ galaxies in one
cluster drawn from the EDisCS sample, at $z = 0.54$. The time interval between
this redshift (when clusters already contained numerous faint red galaxies) and
$z\sim 0.8$ (when clusters exhibit a clear deficit of this population) is
approximately $1.5$~Gyr, comparable with the typical time-scale probed by a k+a
spectrum. Below, we complement these spectroscopic data with EDisCS data
available for the cluster cl1138.2-1133 (hereafter cl1138). This cluster was
originally targeted as a high-z ($z_{\rm est}\sim 0.8$) cluster candidate, but
it turned out that the field contains two clusters at $z\sim 0.5$. For this
cluster, the available EDisCS spectroscopy is as deep as the new spectroscopy
presented here.

The layout of the paper is as follows. Our target selection, observations, and
data reduction are described in Section \ref{sec:data}. In Section
\ref{sec:specanalysis}, we characterise the spectral composition of the
clusters used in this study. In Section \ref{sec:ratio}, we discuss our new
results in the context of the build-up of the red-sequence galaxy
population. Finally, we discuss our results and give our conclusions in Section
\ref{sec:discconcl}. Where necessary in the following, we assume a $\Lambda$CDM
cosmology with $\Omega_{\Lambda} = 0.7$, $\Omega_{\rm m} = 0.3$, and H$_0 =
70\, {\rm km} {\rm s}^{-1}{\rm Mpc}^{-1}$.

\section{The data}
\label{sec:data}

Our target cluster for new spectroscopic observations (cl1232.5-1250, hereafter
abbreviated as cl1232) is a rich cluster at $z=0.54$ with a line-of-sight
velocity dispersion of $1080^{+119}_{-89}\,{\rm km}\,{\rm s^{-1}}$
\citep{Halliday_etal_2004}, and a significant weak lensing detection
\citep{Clowe_etal_2006}. For this cluster, the EDisCS programme obtained deep
FORS2 \citep{Appenzeller_etal_1998} photometry at $V$, $R$, and $I$ with
integration times of approximately 2 hours \citep{White_etal_2005}, and deep
near infrared imaging using SOFI at NTT, with integration times of 300 min at
$J$ and 360 min at $K_s$ (Arag\'on-Salamanca et al. in preparation). Within the
EDisCS programme, we also obtained for this cluster deep spectroscopy using 4
separate slit masks on FORS2 and exposure times of 106 (for one mask) and 120
(for the other three masks) minutes \citep{Halliday_etal_2004}. For cl1232, we
also have mosaic images in the F814W filter from the Advanced Camera for
Surveys (ACS) onboard the Hubble Space Telescope (HST), with one orbit exposure
in the outer regions, and five orbit coverage of the cluster centre
\citep{Desai_etal_2007}.

For our new spectroscopic observations, target selection was based on the
available VLT/FORS2 photometry \citep{White_etal_2005}. The optical data cover
$6.5\arcmin \times 6.5\arcmin$ and are well matched to the FORS2 spectrograph
field of view. Our targets were selected among galaxies with I-band magnitude
between 21.5 and 23. We excluded galaxies with V-I colour redder than 2.6, so as
to exclude targets above the observed colour-magnitude relation for this
cluster. At fainter magnitudes, when possible, we added slits on targets at
$23<$~I~$<24$ within a wedge-shaped region close to the observed
colour-magnitude relation (see Fig.~\ref{fig:cm} below). For the target
selection, we used magnitudes and colours computed within an aperture of radius
$1\arcsec$, and uncorrected for Galactic extinction. The spectroscopic data
presented in this work are deeper than available EDisCS spectroscopy by about
$1$ magnitude.

Spectroscopic data were acquired during one observing run spanning four nights
(starting on April 16, 2004) using the MXU multi-object mask facility of the
FORS2 spectrograph, mounted on the VLT Yepun UT4 telescope in Paranal. The
planned observations could not be carried out due to bad weather conditions
(strong wind, high humidity, and seeing always above $1.1\arcsec$ and varying
strongly, up to $1.7\arcsec$ during integration). Observations for only three
masks out of the eight prepared were completed, with exposure times of 4.5
hours and using the high efficiency grism 600RI+19 ($\lambda_{\rm central} =
6780$~\AA, resolution ${\rm FWHM} \approx 6$~\AA). The signal-to-noise ratio
around $6500$~\AA ~is $\sim 9.6$ at I=21.6, $\sim 2.8$ at I=22.4, and $\sim 2$
per FORS2 pixel ($1.66$~\AA/pixel) at I=23.0. Because of poor seeing, the
quality of these spectra is too low to perform a detailed spectroscopic
analysis. Nevertheless, as we discuss in the following, these data have been
useful in assessing our strategy and quantifying the role of the post-starburst
phase in the build up of the red-sequence for the cluster under investigation.

Data reduction was completed using IRAF\footnote{IRAF is distributed by the
  National Optical Astronomy Observatories, which are operated by the
  Association of Universities for Research in Astronomy, Inc., under
  cooperative agreement with the National Science Fundation. See:
  http://iraf.noao.edu/} and the improved sky subtraction method, flux
calibration, and correction for telluric absorption, which are described in
detail in \citet[][see also \citealt{Kelson_2003} and
  \citealt{Halliday_etal_2004}]{Milvang-Jensen_etal_2008}.  We refer to these
papers for full details of the data reduction procedure.

Spectroscopic redshifts were measured using emission lines where possible, in
particular the [OII]3727 line, or the most prominent absorption lines,
e.g. calcium K and H lines at 3934\AA ~and 3968\AA. A quality flag was assigned
manually to each measurement. In the following, we use only `secure' redshift
measurements. The equivalent widths of the [OII] and Balmer lines were measured
with a line fitting technique similar to that used in
\citet{Dressler_etal_1999}, after inspecting each one-dimensional spectrum
interactively. Each two-dimensional spectrum was also inspected so as to
confirm the presence of an emission line, particularly in the case of weak
lines.

As explained in Section~\ref{sec:intro}, cl1138 was originally targeted as a
high-$z$ cluster, with an expected redshift $z_{\rm est}\sim 0.8$. The redshift
of all candidate clusters observed within the EDisCS programme was estimated
using an empirically calibrated relation between the redshift and the magnitude
of the brightest galaxies \citep{Gonzalez_etal_2001}. For this cluster,
however, a large part of the imaging data near the cluster detection was masked
(likely due to a nearby bright star). As a result, the cluster centroid was
shifted slightly away from the actual brightest cluster galaxy (BCG), which
ended up just outside the automated search radius. As a result, a fainter
galaxy was identified as BCG and the cluster redshift was overestimated
(Gonzalez, private communication). The dominant spectroscopic peak for cl1138
is at $z=0.48$ (45 secure spectroscopic members). The field also contains a
secondary peak at $z=0.45$, with 11 secure spectroscopic members
\citep{Milvang-Jensen_etal_2008}. The main cluster in the field has a
line-of-sight velocity dispersion of $732^{+72}_{-76}\,{\rm km}\,{\rm s}^{-1}$,
while the secondary cluster has a velocity dispersion of $542^{+63}_{-71}\,{\rm
  km}\,{\rm s}^{-1}$. Since the field was targeted as a high-$z$ cluster
candidate, the available spectroscopy is deeper than that obtained for other
EDisCS clusters at similar redshift, and extends down to a limiting magnitude
similar to that obtained for cl1232 with our new spectroscopic
observations. The spectroscopic data reduction and spectral analysis for this
cluster was carried out using the same methods and procedures outlined
above. We note that our photometric redshift estimates are not robust at
$z\lesssim 0.5$ for the filter set used for the high-redshift candidate
clusters such as cl1138 (Pell\'o et al. submitted). B-band data would be needed
for this cluster to achieve higher accuracy in the photometric redshift
estimation.

\begin{table*}
\caption{EDisCS ID, right ascension, declination, redshift, EWs of the OII and
  H$\zeta$ lines, and spectral type for all spectroscopic members provided by
  the data presented in this study. Members already observed within the EDisCS
  programme are indicated with an asterisk. Unmeasurable values of EWs are
  indicated by '---'.}

\begin{tabular}{lllllll}
  \hline
EDisCS ID & RA (J2000) & Dec (J2000) & z & EW[OII] & EW[H$\zeta$] & spectral type\\
\hline
EDCSNJ1232370-1248239(*) & 12:32:37.04 & -12:48:23.94 & 0.5399 & 0.0 & 2.7 & passive\\
EDCSNJ1232348-1248427 & 12:32:34.76 & -12:48:42.73  &  0.5335 & 80.3 & 0.0 & active \\
EDCSNJ1232342-1248565 & 12:32:34.20 & -12:48:56.45  &  0.5323 & 44.2 & 0.0 & active \\
EDCSNJ1232311-1249392 & 12:32:31.11 & -12:49:39.16  &  0.5445 & --- & --- & uncertain\\
EDCSNJ1232350-1249519 & 12:32:34.98 & -12:49:51.91  &  0.5450 & --- & --- & uncertain\\
EDCSNJ1232276-1250017 & 12:32:27.61 & -12:50:01.69  &  0.5411 & 0.0 & 0.0 & passive \\
EDCSNJ1232274-1250548 & 12:32:27.43 & -12:50:54.78  &  0.5441 & 0.0 & 0.0 & passive \\
EDCSNJ1232291-1251054 & 12:32:29.08 & -12:51:05.40  &  0.5433 & 0.0 & 1.3 & passive\\
EDCSNJ1232212-1251476 & 12:32:21.16 & -12:51:47.61  &  0.5399 & 0.0 & --- & passive\\
EDCSNJ1232284-1251573 & 12:32:28.36 & -12:51:57.26  &  0.5437 & 0.0 & 0.0 & passive \\
EDCSNJ1232242-1252301(*) & 12:32:24.24 & -12:52:30.10 & 0.5386 & 0.0 & 2.9 & passive\\
EDCSNJ1232261-1252429(*) & 12:32:26.11 & -12:52:42.94 & 0.5376 & 0.0 & 3.0 & (weak) k+a\\
EDCSNJ1232230-1253041 & 12:32:22.97 & -12:53:04.07  &  0.5448 & 0.0 & 0.0 & passive \\
EDCSNJ1232369-1248246 & 12:32:36.88 & -12:48:24.58  &  0.5362 &  7.0 & 0.0 & active \\
EDCSNJ1232366-1248446 & 12:32:36.61 & -12:48:44.62  &  0.5352 & 53.3 & 0.0 & active \\
EDCSNJ1232335-1249293 & 12:32:33.54 & -12:49:29.29  &  0.5450 & 37.4 & 8.0 & active \\
EDCSNJ1232292-1249417 & 12:32:29.17 & -12:49:41.70  &  0.5360 & --- & ---  & uncertain\\
EDCSNJ1232299-1250018 & 12:32:29.94 & -12:50:01.84  &  0.5409 & 0.0 & 0.0 & passive \\
EDCSNJ1232339-1250359 & 12:32:33.89 & -12:50:35.93  &  0.5416 & 0.0 & 0.0 & passive \\
EDCSNJ1232285-1250527 & 12:32:28.53 & -12:50:52.74  &  0.5416 & 0.0 & 0.0 & passive \\
EDCSNJ1232300-1251293 & 12:32:29.97 & -12:51:29.30  &  0.5419 & --- & --- & uncertain\\
EDCSNJ1232333-1252436 & 12:32:33.32 & -12:52:43.56  &  0.5430 &  9.3 &  2.7 & active\\
EDCSNJ1232320-1252551 & 12:32:31.97 & -12:52:55.15  &  0.5391 & 26.4 & 0.0 & active \\
EDCSNJ1232358-1249348 & 12:32:35.80 & -12:49:34.77  &  0.5328 & 0.0 & 0.0 & passive \\
EDCSNJ1232319-1250024 & 12:32:31.91 & -12:50:02.44  &  0.5475 & --- & --- & uncertain\\
EDCSNJ1232366-1250526 & 12:32:36.61 & -12:50:52.63  &  0.5391 & 0.0 & 0.0 & passive \\
EDCSNJ1232320-1251257 & 12:32:32.04 & -12:51:25.72  &  0.5485 & 41.1 & 0.0 & active \\
EDCSNJ1232311-1251563 & 12:32:31.07 & -12:51:56.32  &  0.5407 & 0.0 & 0.0 & passive\\
EDCSNJ1232261-1253040(*) & 12:32:26.08 & -12:53:04.00 & 0.5597 & 5.0 & 0.0 & active \\
EDCSNJ1232324-1253496 & 12:32:32.44 & -12:53:49.57  &  0.5407 & 0.0 & 11.4 & k+a\\
EDCSNJ1232361-1254185(*) & 12:32:36.10 & -12:54:18.52 & 0.5398 & 2.3 & 6.5 & k+a\\

\hline
\end{tabular}
\label{tab:specinfo}
\end{table*}

\section{Spectral composition}
\label{sec:specanalysis}

\begin{figure*}
\bc
\hspace{-0.3truecm}
\resizebox{18.cm}{!}{\includegraphics[]{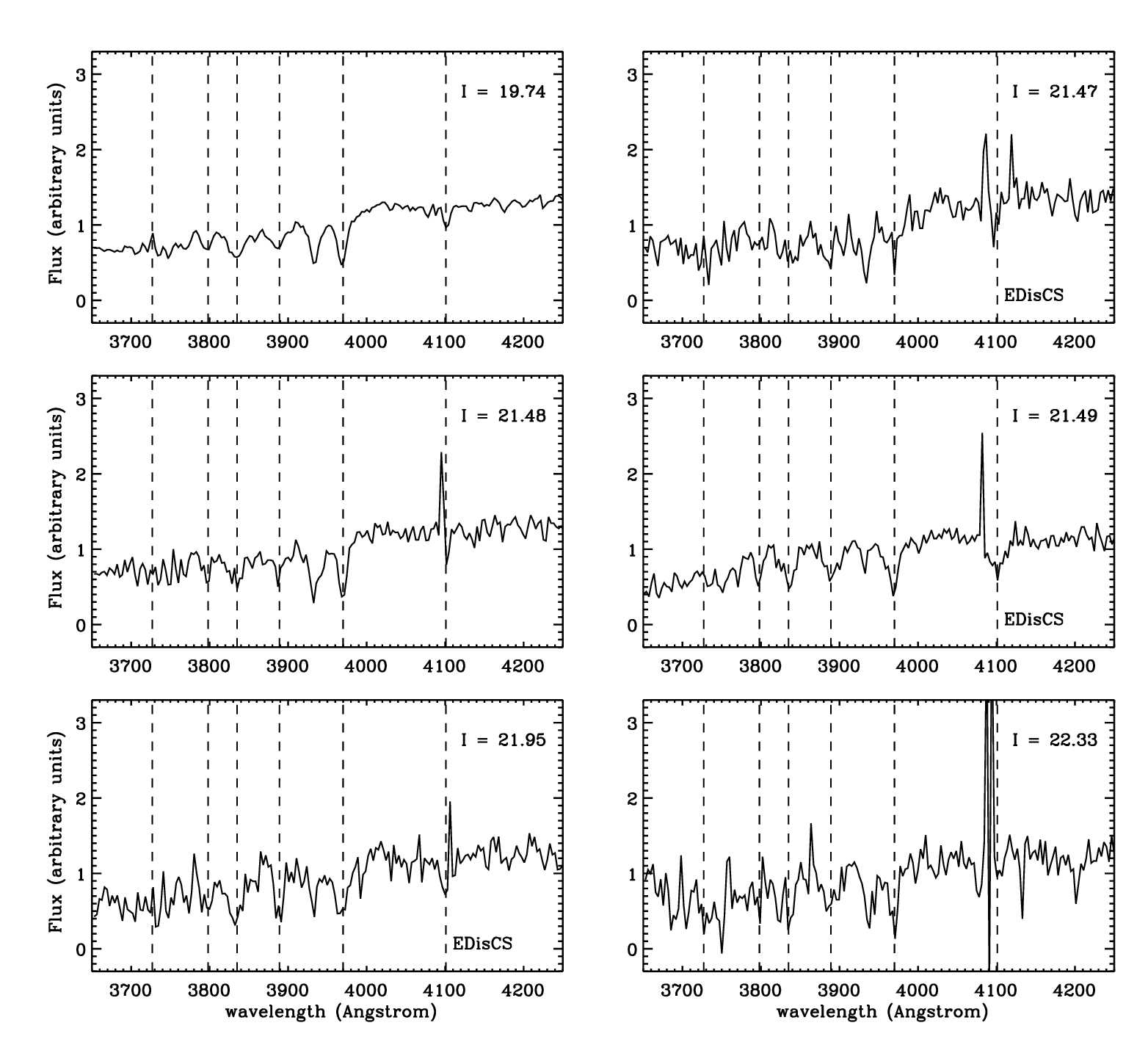}}
\caption{Spectra of the cl1232 members classified as post-starburst galaxies,
  in order of increasing luminosity. Dashed vertical lines mark from left to
  right the following lines: [OII]3727, H$\theta$3798, H$\eta$3835,
  H$\zeta$3889, H$\epsilon$3970, and H$\delta$4101. Spectra are shown in the
  rest frame and have been rebinned at 5~\AA.}
\label{fig:spectra}
\ec
\end{figure*}

The new spectroscopic data provide 31 cluster members for the cluster
cl1232. Five of these members were already observed within the EDisCS
programme, yielding a total of 26 new cluster members. The additional
spectroscopic information does not modify significantly the dynamical and
substructure analysis of this cluster discussed in
\citet{Halliday_etal_2004}. In the following, we therefore focus on an analysis
of the spectral composition of the cluster cl1232, and compare it to the
information available for cl1138.

We grouped all spectroscopic members into three broad spectral classes:
galaxies with an absorption-line spectrum are classified as \emph{passive},
while galaxies with significant emission in their spectra (EW[OII]$> 5$~\AA)
are classified as \emph{active}. In addition, we classify as k+a galaxies all
those with no significant [OII] emission ($< 5$~\AA), and with strong Balmer
lines in absorption. At the redshift of cl1232, unfortunately, the 6300 \AA
~sky line falls exactly on top of the H$\delta$ line. For our k+a
classification, we therefore rely on higher-order Balmer lines, in particular
H$\zeta$, which is strongly correlated to H$\delta$, and provides a useful
substitute whenever H$\delta$ is not available \citep{Poggianti_etal_2009}. In
our study, we classify as post-starburst galaxies all those with EW$[{\rm
    H}\zeta]\ge 3$~\AA~and EW[OII]$<5$~\AA.

In Table~\ref{tab:specinfo}, we provide, for each cluster member obtained
through our new observations, the EDisCS object ID (column 1), the right
ascension (column 2), the declination (column 3), the measured redshift (column
4), the equivalent widths of the [OII] (column 5) and H$\zeta$ (column 6)
lines, and the spectral type (column 7). Members already observed in the EDisCS
programme are indicated with an asterisk.

The spectra of the six members classified as post-starburst galaxies are shown
in Fig.~\ref{fig:spectra}, in order of increasing luminosity. The dashed
vertical lines in the figure mark the main spectral lines over the wavelength
range shown: [OII]3727, H$\theta$3798, H$\eta$3835, H$\zeta$3889,
H$\epsilon$3970, and H$\delta$4101. Spectra are shown in the rest frame and
have been rebinned to 5~\AA/pixel. The signal-to-noise ratio decreases towards
fainter magnitudes and, as discussed above, a sky line at 6300~\AA~is clearly
visible in most of the panels and complicates significantly the measurement of
the H$\delta$ equivalent width.

\begin{figure*}
\bc
\hspace{-0.3truecm}
\resizebox{18.cm}{!}{\includegraphics[]{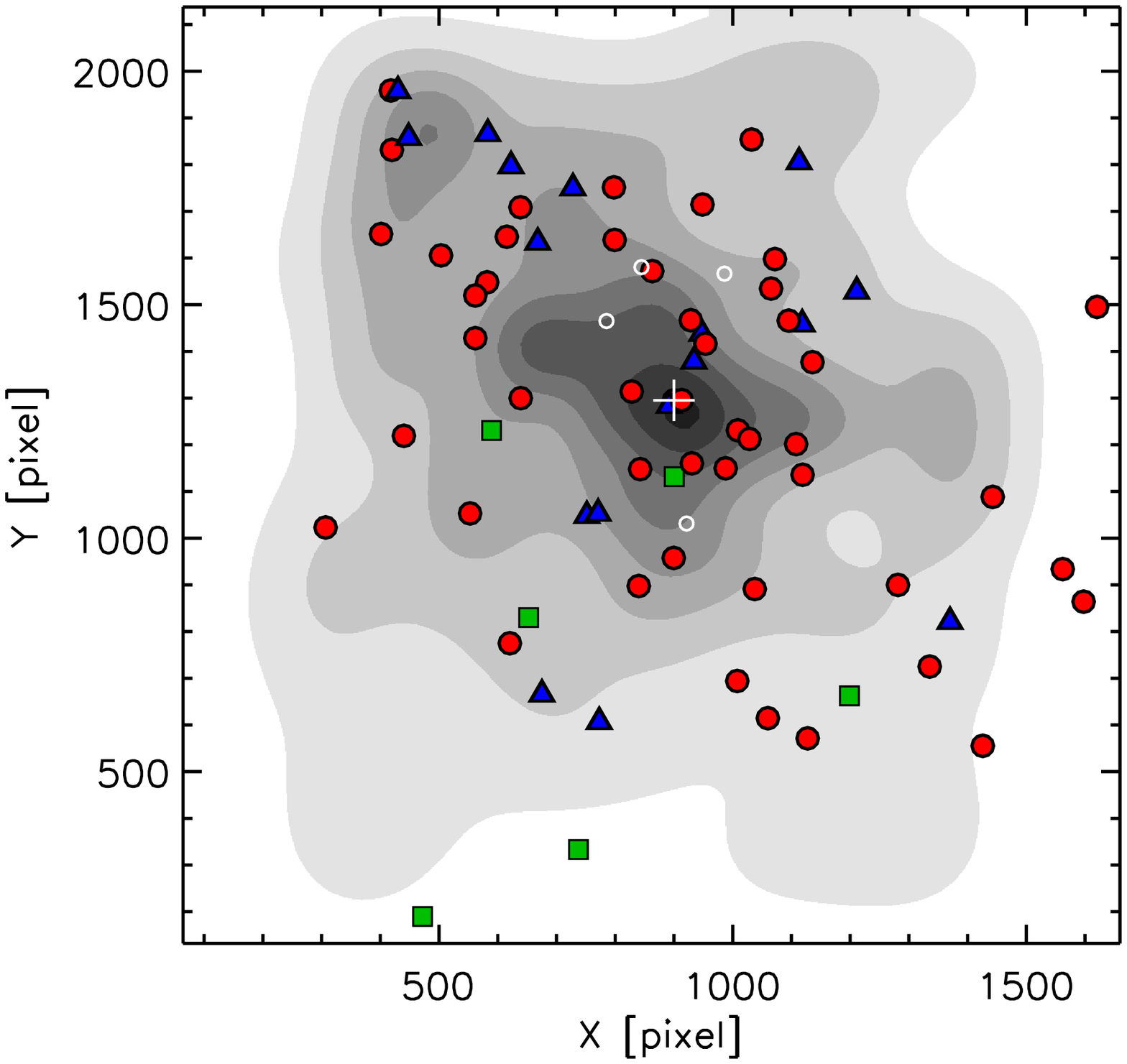}
                     \includegraphics[]{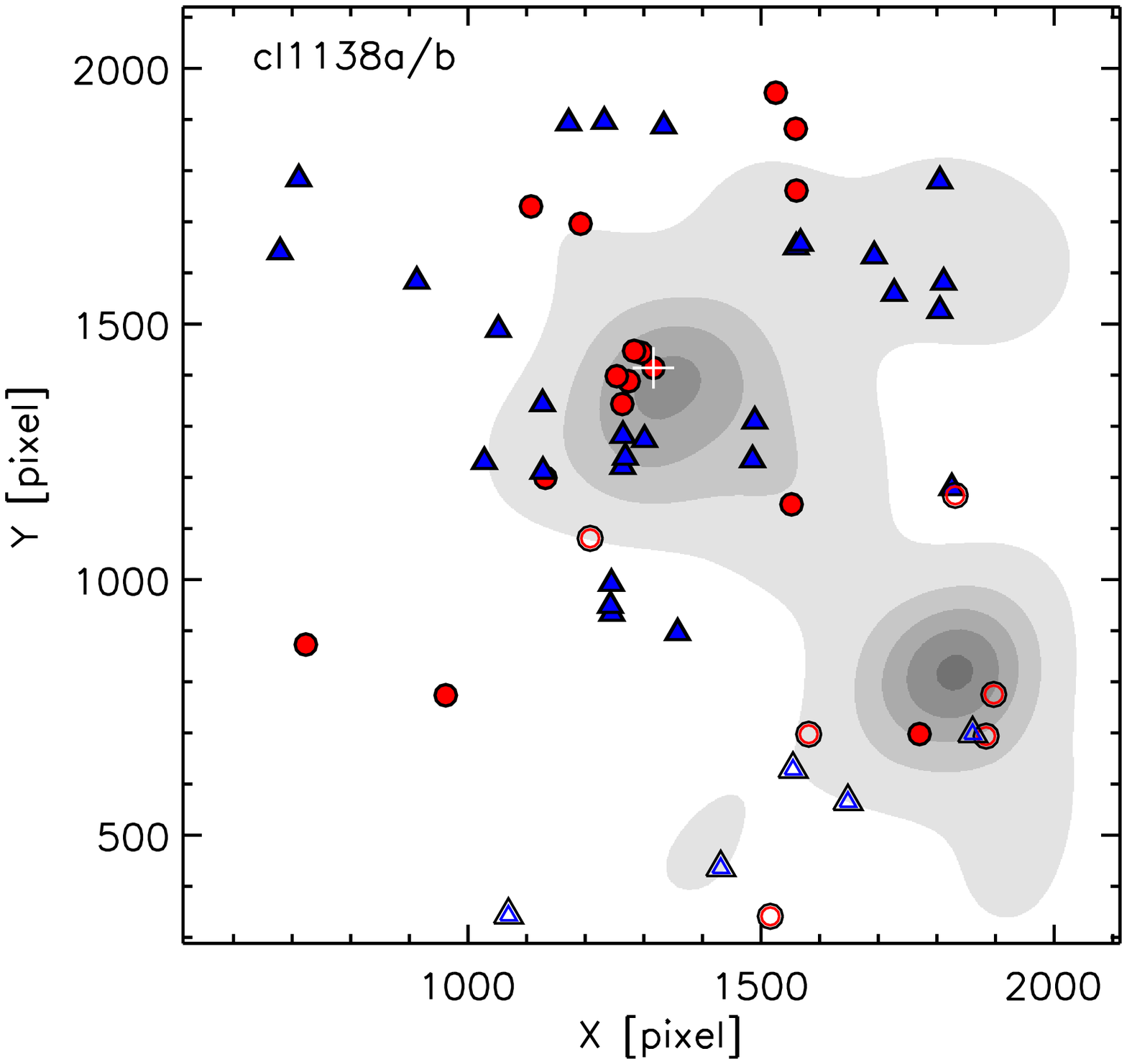}}
\caption{Spatial distribution of all spectroscopic members for the clusters
  cl1232 (left panel) and cl1138 (right panel), superimposed on adaptively
  smoothed isopleths of the surface density of photo-z members. Red circles and
  blue triangles correspond to galaxies with absorption and emission-line
  spectra, respectively. Green boxes mark the position of galaxies classified
  as k+a. Galaxies with uncertain spectroscopic type are shown as empty white
  circles. White crosses mark the position of the BCG of the main spectroscopic
  peak in each field. In the right panel, spectroscopic members of the main
  cluster are plotted as filled symbols, while empty symbols represent the
  spectroscopic members of the secondary cluster in the field. The pixel scale
  is of $0.2\arcsec$/pixel.}
\label{fig:spatialdistr}
\ec
\end{figure*}

Fig.~\ref{fig:spatialdistr} shows the spatial distribution of all cluster
members for cl1232 (left panel) and cl1138 (right panel), superimposed on
adaptively smoothed isopleths created using all photometric redshift members in
the region where both optical and near infrared data are available, and a
scheme very similar to that outlined in \citet{Pisani_1993,Pisani_1996}. Red
circles and blue triangles mark the position of passive and active galaxies
respectively, while green boxes correspond to galaxies classified as k+a. A few
galaxies could not be assigned a `secure' spectroscopic
type\footnote{E.g. because the measurement of line intensities is contaminated
  by the presence of sky lines} and are shown as open white circles in
Fig.~\ref{fig:spatialdistr}. The adaptively smoothed maps were constructed
using the locations of all galaxies brighter than $I(r=1\arcsec) = 25$ that
have a high probability of being cluster members, according to the redshift
probability distribution provided by two different photometric redshift codes
(see section 3.1 in \citealt{DeLucia_etal_2007} and Pell\'o et
al. submitted). Note that we have not attempted to correct for boundary
effects, so the estimated densities always drop close to the field boundaries.

Cl1232 is clearly dominated by a population of passive galaxies (about 62 per
cent of all spectroscopic members) concentrated around the position of the BCG,
which is marked by a white cross in Fig.~\ref{fig:spatialdistr}. About 25 per
cent of all members have an emission-line spectrum. These galaxies appear more
uniformly distributed, with a small concentration in the north-west corner of
the image, which might correspond to a small sub-clump infalling onto the main
body of the cluster (this is confirmed by a Dressler-Shectman test on the
available spectroscopy). Among all spectroscopic members of cl1232, we find 6
that are classified as post-starburst galaxies. Two of these are in common
between the EDisCS sample and the sample presented in this study. One of them
was classified as `passive' on the basis of the available EDisCS spectroscopy,
but is classified as k+a according to our deeper spectra. Most of the k+a
galaxies are located in the southern region of the cluster and do not appear to
reside preferentially in regions of high density. Interestingly, two of these
galaxies are located in the south-west corner of the image, where our weak
lensing reconstruction detects a second peak at more than $3\sigma$
significance \citep{Clowe_etal_2006}. This could suggest that these k+a
galaxies originate in infalling groups from e.g. interactions with the
intra-cluster medium. Unfortunately, however, these two galaxies are located
very close to the field boundary, where our density reconstruction is not
accurate.

The right panel of Fig.~\ref{fig:spatialdistr} shows the spatial distribution
of spectroscopic members for both clusters identified in the field of cl1138,
and the corresponding isopleths. Filled symbols represent the spectroscopic
members of the main cluster, while open symbols correspond to the secondary
clump in this field. The spectral properties of cl1138 are quite different from
those of cl1232. Our EDisCS spectroscopy does not include any galaxy classified
as post-starburst, and the star-forming galaxy fraction is twice as large in
cl1138 as it is in cl1232 \citep{Poggianti_etal_2006}.  The photometric
redshift uncertainty does not allow us to isolate members of the two clusters
in the field of cl1138, so that both of them appear in the corresponding
isopleths.

\section{Colour-magnitude relation}
\label{sec:ratio}

In \citet{DeLucia_etal_2007}, we analysed the colour-magnitude relation of all
clusters in the EDisCS sample and showed, confirming our previous findings,
that the high redshift EDisCS clusters exhibit a significant \emph{deficit} of
faint red galaxies. Our results were later confirmed by an independent analysis
of the red-sequence luminosity function of the EDisCS clusters
\citep{Rudnick_etal_2009}. As mentioned in Section \ref{sec:intro}, we
interpret these results in terms of a progressive build-up of the faint end of
the red-sequence with decreasing redshift, possibly due to star formation being
suppressed in infalling galaxies by the hostile cluster environment. In this
section, we revisit our results for cl1232 using the new spectroscopic data
presented in this study, and reanalyse our results for cl1138.

\begin{figure}
\bc
\hspace{-0.3truecm}
\resizebox{8.6cm}{!}{\includegraphics[]{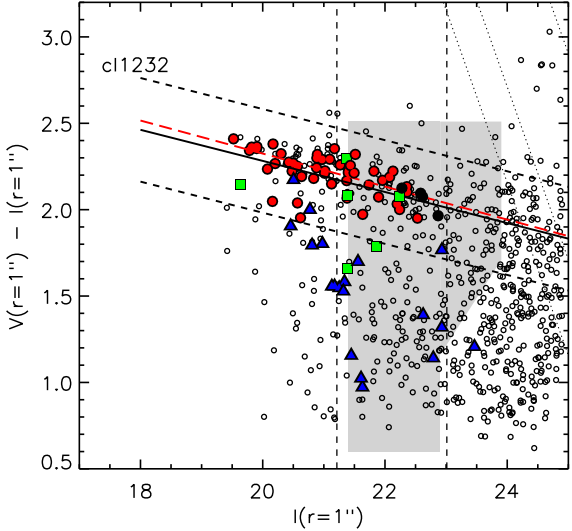}}
\caption{Colour-magnitude relation for the cluster cl1232. Black open symbols
  represent photo-z cluster members, while blue and red symbols correspond to
  spectroscopically confirmed members with and without emission lines. Filled
  black symbols correspond to spectroscopic members with uncertain
  spectroscopic type, and green squares show galaxies that have been classified
  as post-starburst. Thin slanted lines correspond to the $1$, $3$, and
  $5\sigma$ detection limit in the V-band. The solid black line shows the best
  fit relation used in \citet{DeLucia_etal_2007}, while the dashed red line
  shows the best fit relation re-measured using all cluster members (see
  text). The grey shaded region mark the region used for our target selection
  (see Section~\ref{sec:data}), with limits corrected for Galactic extinction
  as done for the photometric data shown in the figure. The vertical dashed
  lines mark the limits chosen to classify galaxies as `faint' and `luminous'
  (see text for details).}
\label{fig:cm}
\ec
\end{figure}

Fig.~\ref{fig:cm} shows the colour-magnitude diagram for the cluster
cl1232. Small black open circles show galaxies for which the photometric
redshift measurements provide a high probability of cluster membership, while
larger symbols show spectroscopically confirmed members with (blue triangles)
and without emission lines (red circles), and with a post-starburst spectrum
(green boxes). The few spectroscopic members with uncertain spectral type are
shown as filled black circles, and lie almost exactly on the faint end of the
red-sequence for this cluster. The solid black line in Fig.~\ref{fig:cm} shows
the best-fit relation to the red-sequence used in \citet{DeLucia_etal_2007},
which was obtained by applying the bi-weight estimator \citep*{Beers_etal_1990}
for the members without emission lines in their spectra, and adopting a fixed
slope of $-0.09$. The grey shaded region corresponds to the region used for our
target selection. As in \citet{DeLucia_etal_2007}, we used magnitudes and
colours measured within a fixed circular aperture with $1\arcsec$ radius,
corrected for Galactic extinction.

The data shown in Fig.~\ref{fig:cm} demonstrate that the best-fit relation used
in \citet{DeLucia_etal_2007} provides a good fit to the observed red-sequence,
down to the magnitude limit sampled by the new data presented in this study (we
recall that the EDisCS spectroscopy extends down to ~22 in the I-band for
this cluster - see Fig.~1 in De Lucia et al.~2007). The dashed red line in
Fig.~\ref{fig:cm} shows the best-fit relation obtained using the additional
spectroscopic information presented in this work, and leaving both the
zero-point and the slope of the relation free to vary. In this case, we derive a
best-fit slope of $-0.095\pm0.017$, in very good agreement with the fixed slope
adopted in our previous study. 

Our fitting procedure measures a scatter of $0.082\pm0.014$. Interestingly,
this scatter is comparable to the scatter measured for a sample of clusters at
similar redshift, observed with HST by \citet{Ellis_etal_1997}\footnote{Note
  that \citet{Ellis_etal_1997} estimated the rms scatter about the mean
  regression from the mean absolute deviation, which would generally yield a
  lower value than that provided by the method adopted in this
  study.}. Uncertainties in the fit coefficients given above were estimated by
bootstrapping the available data from 100 Monte Carlo simulations.

As noted in the previous section, cl1232 has a substantial population of
passive galaxies that lie on a relatively tight red-sequence.  The
spectroscopic members with emission lines in their spectrum scatter below the
best-fit relation, with a few of them having colours compatible with those of
members with an absorption-line spectrum. Of the six members that have been
classified as post-starburst galaxies, four of them have V-I colour within $\pm
0.3$~mag from the best-fit relation (dashed lines), and two are slightly
bluer. Our data seem to exclude a large (larger than the number obtained on the
basis of the EDisCS spectroscopy for brighter galaxies) population of
post-starburst galaxies at faint magnitudes ($M_V > -18.5$), as found for the
Coma cluster \citep{Poggianti_etal_2004}.

\begin{figure}
\bc
\hspace{-0.3truecm}
\resizebox{8.6cm}{!}{\includegraphics[]{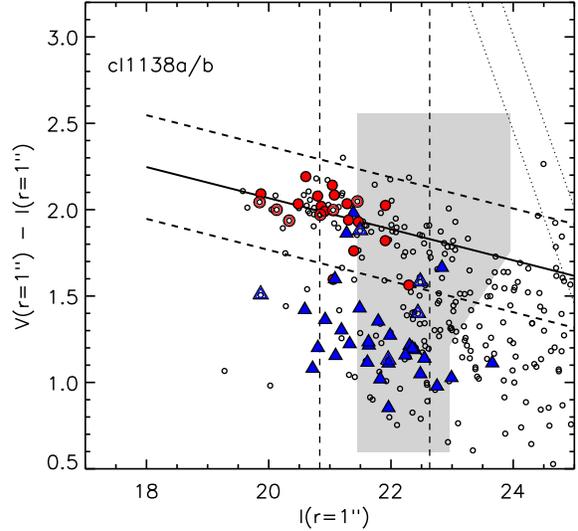}}
\caption{As in Fig.~\ref{fig:cm} but for the two clusters in the field of
  cl1138. Filled symbols show the spectroscopic members of the main cluster,
  while open symbols correspond to the secondary cluster in this field. All
  members for this clusters come from EDisCS spectroscopy (see text).}
\label{fig:cm1138}
\ec
\end{figure}

As discussed in Sec.~\ref{sec:specanalysis}, there is a large variation in the
spectral composition of clusters at similar redshift. It is therefore
interesting to compare the colour-magnitude relation of the cl1232 cluster to
that of cl1138. This is shown in Fig.~\ref{fig:cm1138}, using the same
colour-coding adopted in Fig.~\ref{fig:cm}. Spectroscopic members of the
secondary cluster in the field of cl1138 are shown as empty symbols. The black
solid line shows the best-fit relation used in \citet{DeLucia_etal_2007}, while
the dashed lines correspond to $\pm 0.3$~mag from it. For this cluster, a fit
of the red-sequence using only members with an absorption line spectrum does
not converge when leaving both the slope and the zero-point of the relation
free to vary. Fig.~\ref{fig:cm1138} shows a tight red-sequence at bright
magnitudes, with few passive members scattering below the best-fit relation at
fainter magnitude. The population of active galaxies, more copious than for the
cluster cl1232, is characterised by significantly blue colours, although a few
members with emission lines in their spectra reside within $0.3$~mag from the
best-fit red-sequence.

Figs.~\ref{fig:cm} and \ref{fig:cm1138} show that the relative proportions of
red and blue galaxies are quite different for the two clusters under
consideration. In addition, as noted in the previous section, no spectroscopic
member of cl1138 has been classified as `post-starburst'.

\begin{figure}
\bc
\hspace{-0.3truecm}
\resizebox{8.6cm}{!}{\includegraphics[]{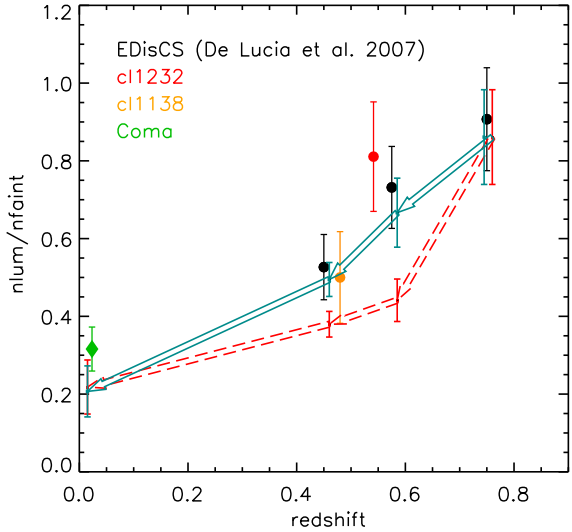}}
\caption{Luminous-to-faint ratio for cl1232 (red circle) and cl1138 (orange
  circle), compared to the average trend obtained from the EDisCS clusters
  (black circles), as measured in \citet{DeLucia_etal_2007}. The green diamond
  shows the value measured for the Coma cluster in the same study. The red and
  cyan error bars connected by arrows show the evolution of the
  luminous-to-faint ratio expected using the `truncation' and the `delay'
  models described in \citet{DeLucia_etal_2007}.}
\label{fig:ratio}
\ec
\end{figure}

To analyse the distribution of red-sequence galaxies, we measure the
`luminous-to-faint' ratio, following the same definitions adopted in
\citet{DeLucia_etal_2007}. We use all galaxies that lie within $\pm 0.3$~mag
from the best-fit colour-magnitude relation, measured using a fixed slope of
$-0.09$. We then define as `luminous' all galaxies brighter than $M_{\rm V} =
-20$, and as `faint' all galaxies fainter than this and brighter than $M_{\rm
  V} =-18.2$. The conversion from observed I-band magnitudes to rest-frame
V-band magnitudes has been made by assuming a single burst model with redshift
of formation $z_f = 3$, and takes into account passive evolution to redshift
zero. The luminous-to-faint ratio measured\footnote{The measurement includes
  the new spectroscopic information discussed in this study, but it does not
  differ significantly from that obtained using EDisCS spectroscopy only.} for
cl1232 is shown as a red circle in Fig.~\ref{fig:ratio}, while the orange
circle shows the corresponding measurement for the cluster cl1138. The black
circles with error bars show the average luminous-to-faint ratio measured by
binning 15 EDisCS clusters as discussed in \citet{DeLucia_etal_2007}, and the
green diamond shows the corresponding value measured for the Coma cluster. We
recall that the EDisCS values shown in Fig.~\ref{fig:ratio} have been obtained
by averaging the measurements for different choices of cluster membership
criteria and area. A conservative choice was adopted for the error bars, which
were also averaged rather than combined in quadrature. Since our previous work
demonstrated that results were not affected significantly by the use of
different cluster membership criteria, the values for cl1232 and cl1138 shown
in Fig.~\ref{fig:ratio} have been measured by selecting members using the
photometric redshift information. For cl1232, we have selected all photometric
members within the maximum physical radius centred on the BCG and included in
the SOFI field of view. For cl1138, we have selected the photometric members of
both clusters in the field, within the SOFI field of view. The corresponding
value of the luminous-to-faint ratio needs to be interpreted with caution.

Fig.~\ref{fig:ratio} shows that the luminous-to-faint ratios of cl1232 and
cl1138 are consistent, within the errors, with the average trend obtained
considering all EDisCS clusters used in our previous study. In
Fig.~\ref{fig:ratio}, we also show the evolution of the luminous-to-faint ratio
expected using the two models described in \citet{DeLucia_etal_2007}. We refer
the reader to the original paper for details on the models. Briefly, these are
constructed using the colour-magnitude distribution of the EDisCS clusters in
the highest redshift bin, and evolving them forward in time assuming that the
star formation histories of blue galaxies are truncated (`truncation' model) at
the redshift of the observation of the cluster, or 1~Gyr later (`delay'
model). An exponentially declining star formation history (with $\tau = 1$,
$2$, $3$, and $7$~Gyr and redshift of formation $3$) is assigned to all
galaxies bluer by $0.3$~mag than the best-fit red-sequence, by selecting the
``closest model'' in the colour-magnitude space. The adopted modelling is very
simple and, as discussed in detail in \citet{DeLucia_etal_2007}, results need
to be interpreted with caution. Fig.~\ref{fig:ratio} suggests that a scenario
in which infalling galaxies have their star formation suppressed by the cluster
environment is in qualitative agreement with the observed build-up of the
colour-magnitude relation. Given the uncertainties in the modelling, however,
the results discussed in our previous study and here do not yet convincingly
confirm this scenario. In addition, they do not provide strong constraints on
the physical process(es) responsible for the observed evolution, and on the
associated time-scale(s). In the next section, we examine one specific scenario
in which new additions to the colour-magnitude relation move from the blue
cloud passing trough a k+a phase.

\section{The importance of the k+a phase in the build-up of the red-sequence 
  galaxy population}
\label{sec:kplusa}

With the data in hand, we now attempt a first estimate of the contribution to
the build-up of the colour-magnitude relation by galaxies passing through the
k+a phase. Using the definitions given above, and complementing the photometric
redshift information with the available spectroscopy to assign cluster
membership, we have the following numbers of \emph{red} luminous and faint
galaxies for cl1232:

\begin{displaymath}
N_{\rm lum}(z=0.54) = 60 \,\, {\rm and} \,\, N_{\rm faint}(z=0.54) = 74,
\end{displaymath}  
which implies a luminous-to-faint ratio of $\sim 0.81$ at z=0.54.

If $t$ is the age of the Universe corresponding to the redshift of the
cluster\footnote{The age of the Universe corresponding to z=0.54 is $\sim
  8.15$~Gyr.}, the expected number of new red galaxies at a subsequent time
$t+\Delta t$, associated with the k+a phase, can be written as:

\begin{displaymath}
  \Delta N_{\rm lum} (t + \Delta t) = N_{\rm k+a, lum} \times \frac{N_{\rm
      lum}}{N_{\rm lum, spec}}(t) \times \frac{\Delta t}{\tau_{\rm k+a}}, 
\end{displaymath}
where $N_{\rm k+a}$ is the number of {\it blue} k+a galaxies in the magnitude
bin considered. We use here only the blue post-starburst galaxies, because the
red ones have already been included in the computed luminous-to-faint
ratio. For consistency, we use the typical time-scale over which a k+a galaxy
is blue ($\tau_{\rm k+a}$). In the above equation, $N_{\rm lum}$ is the number
of photometric members, and $N_{\rm lum, spec}$ is the number of luminous
spectroscopic members in the same magnitude bin.  The factor $\frac{N_{\rm
    lum}}{N_{\rm lum, spec}}$ corrects for the incompleteness of our
spectroscopic sample, assuming that the sample is representative. We note that
the number of (blue) photometric members is quite uncertain (see
\citealt{DeLucia_etal_2007} and \citealt{Rudnick_etal_2009}) so that these
numbers should be interpreted with caution. A similar equation can be written
for the new additions to the red `faint' galaxies passing through a k+a phase:

\begin{displaymath}
  \Delta N_{\rm faint} (t + \Delta t) = N_{\rm k+a, faint} \times \frac{N_{\rm
      faint}}{N_{\rm faint, spec}}(t) \times \frac{\Delta t}{\tau_{\rm k+a}} 
\end{displaymath}

Assuming $\tau_{\rm k+a} \sim 0.3$~Gyr \citep{Poggianti_Barbaro_1997}, $\Delta
t \sim 0.64$~Gyr, and considering that we observe no blue k+a galaxy in the
`luminous' bin, and two in the `faint' bin, we obtain $\Delta N_{\rm lum}
(z=0.45) = 0.$ and $\Delta N_{\rm faint} (z=0.45)= 13.7$. These numbers give an
expected luminous-to-faint ratio at $z=0.45$ of $\sim 0.68\pm 0.11$, which is
still consistent with the luminous-to-faint ratio measured for cl1232 and
significantly higher than the ratio measured using all (5) EDisCS clusters in
the lower redshift bin used in our previous work (black symbols in
Fig.~\ref{fig:ratio}). When taking into account the (large) estimated
uncertainties, this value becomes marginally consistent with the ratio measured
at lower redshift.

The argument outlined above, however, provides a {\it lower limit} to the
evolution of the luminous-to-faint ratio, because it does not account for the
infall of new galaxies onto the cluster in the redshift interval considered. To
obtain an estimate of the number of galaxies infalling onto a cluster similar
to cl1232 between $z = 0.54$ and $z = 0.45$, we use results from semi-analytic
models of galaxy formation coupled to cosmological N-body simulations. In
particular, we use here the model presented in \citet[][and references
  therein]{DeLucia_Blaizot_2007}, which is publicly available through a
SQL-queryable database\footnote{A description of the publicly available
  catalogues, and a link to the database can be found at the following webpage:
  http://www.mpa-garching.mpg.de/millennium/}. The model has been shown to
provide reasonable agreement with a large number of observational results
both in the local Universe and at higher redshift (see the original paper and
references therein for details).

\begin{figure}
\bc
\hspace{-0.3truecm}
\resizebox{8.6cm}{!}{\includegraphics[]{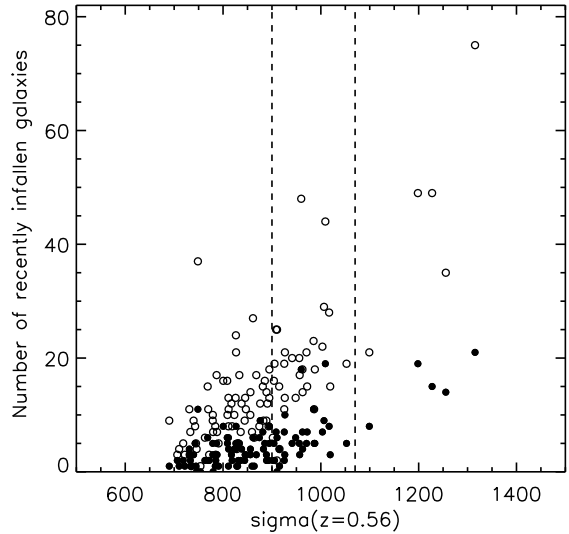}}
\caption{Number of galaxies that fell into the cluster between $z = 0.56$ and
  $z = 0.45$ as a function of the cluster line-of-sight velocity dispersion.
  Only galaxies that are within $0.5\times R_{200}$ from the cluster centre at
  $z = 0.45$ are considered. Filled symbols are obtained counting only galaxies
  brighter than $M_V = -18.6$ at $z = 0.56$, while empty symbols are obtained
  when counting all galaxies brighter than the resolution limit of the
  simulation. The two vertical dashed lines show the region where the
  line-of-sight velocity dispersion is compatible with the value measured for
  cl1232.}
\label{fig:deltan}
\ec
\end{figure}

Using the publicly available catalogues, we selected the 100 most massive
haloes at $z = 0.56$ (the available snapshot closest to the redshift of
cl1232), and identified their descendants at $z = 0.45$. For each of the haloes
at $z = 0.56$, we computed the line-of-sight velocity dispersion using the same
procedure adopted for the EDisCS clusters \citep{Halliday_etal_2004,
  Milvang-Jensen_etal_2008}. We finally counted how many `new' galaxies
(i.e. galaxies which did not reside in the progenitor halo) are found in the
descendant halo at $z = 0.45$. The black filled symbols in
Fig.~\ref{fig:deltan} show the number of recently accreted galaxies that are
found within $0.5\times R_{200}$ from the cluster centre at $z=0.45$, and that
are brighter than $M_{\rm V}=-18.6$ at $z = 0.56$. The adopted distance
corresponds approximately to the maximum fraction of $R_{200}$, for the EDisCS
clusters, that lies within the region covered by the SOFI field of
view. Assuming passive evolution, these galaxies will evolve into galaxies
brighter than $M_{\rm V}=-18.2$ (the magnitude limit adopted in our analysis) at
$z = 0.45$. The dashed vertical lines in Fig.~\ref{fig:deltan} show the range
of velocity dispersions corresponding to the value measured for the cluster
cl1232. To illustrate how these results depend on the adopted magnitude limit,
we show as empty symbols the number of recently accreted galaxies that lie
within $0.5\times R_{200}$ at the lower redshift considered, and that are
brighter than the magnitude limit corresponding to the resolution limit of the
simulation ($\sim -15$ in the rest-frame V-band).

Fig.~\ref{fig:deltan} shows that a halo with velocity dispersion similar to
that measured for cl1232 accretes, on average, $\sim 20$ galaxies brighter than
$M_{\rm V}\sim-15$ in the redshift interval considered, with a few haloes
accreting up to $\sim 50$ new galaxies. The numbers are much lower when
imposing a magnitude cut similar to the magnitude limit adopted in our
analysis: when counting only galaxies brighter than -18.6 at $z = 0.56$, the
average number of newly accreted galaxies is $\sim 6$, and only two haloes in
the relevant range of velocity dispersion accrete $\sim 20$ new galaxies.

Considering that the active and post-starburst galaxies in our sample have been
forming stars at significant rates in the past $\sim 2$~Gyrs, and that a k+a
spectrum requires a sharp truncation of the star formation activity in the past
$\sim 1$~Gyr, the ratio of the number of k+a to the total number of
active and k+a galaxies provides an estimate of the `quenching efficiency'
associated with the particular environment under consideration
\citep[see][]{Poggianti_etal_2009}. Our spectroscopic sample for the cluster
cl1232 contains 6 k+a galaxies and 26 active galaxies (counting also those with
a k+a spectrum), giving a quenching efficiency of $\sim 23$ per cent, in
perfect agreement with the value measured by \citet[][see their
  Table~6]{Poggianti_etal_2009}. We therefore expect $0.23\times N_{\rm accr}$
new red galaxies at $z = 0.45$, which can be added to the numbers computed
above on the basis of the observed number of blue k+a galaxies. If we assume
that all new galaxies accreted onto the cluster end up on the `faint' end of
the colour-magnitude relation - which maximises the evolution of the
luminous-to-faint ratio - we obtain:

\begin{displaymath}
  \frac{N_{\rm lum}}{N_{\rm faint}}(z = 0.45) = \frac{60}{74+13.7+0.23\times
    N_{\rm accr}}  
\end{displaymath}
With $N_{\rm accr} \sim 6$, the above equation gives $\frac{N_{\rm lum}}{N_{\rm
    faint}}(z = 0.45) \sim 0.67\pm 0.11$. Even maximizing the number of newly
accreted galaxies ($N_{\rm accr} \sim 20$), the expected luminous-to-faint
ratio at $z= 0.45$ becomes $\sim 0.65\pm 0.11$, which is not significantly
different from the ratio obtained on the basis of the observed number of blue
k+a galaxies. The number of newly accreted galaxies necessary to bring the
expected luminous-to-faint ratio into agreement with the value measured at
$z\sim0.45$ is $\sim 100$, much larger than the numbers measured for any of the
haloes shown in Fig.~\ref{fig:deltan}, even when counting all galaxies down to
the resolution limit of the simulation.
 
The argument outlined in this section suggests that the observed number of k+a
galaxies in cl1232 is too low to account for the observed progressive build-up
of the faint red-sequence galaxy population. Cl1232 is only one cluster, and we
know that there is a large cluster-to-cluster variation. The only other cluster
for which we have as deep spectroscopy at approximately the same redshift is
cl1138. As explained above, no spectroscopic member of this cluster is
classified as post-starburst, {so no new red galaxy addition to the faint end
  of the red-sequence is expected to pass through a k+a phase for this
  cluster}. On the basis of available data then, k+a galaxies do not seem to
play an important role in the building-up of the colour-magnitude relation of
intermediate redshift clusters. More data are necessary, however, to draw
statistically robust conclusions.

\section{Discussion and conclusions}
\label{sec:discconcl}

We present new spectroscopic observations for a cluster at intermediate
redshift (cl1232 at z=0.54). The cluster belongs to the sample studied by the
ESO Distant Cluster Survey (EDisCS) and the new data presented here extend the
available spectroscopy by about 1 magnitude. Cl1232 has a large fraction of
passive galaxies, and exhibits a well defined and relatively tight
colour-magnitude relation. Among its spectroscopic members, we find six
galaxies that can be classified as `post-starburst' (k+a) systems, on the basis
of the lack of significant emission and strong Balmer lines in absorption.  The
EDisCS sample contains another cluster (cl1138) at similar redshift that was
originally targeted as a high-redshift ($z_{\rm est}\sim 0.8$) cluster, and
whose spectroscopy extends down to similar limiting magnitude as that obtained
for cl1232. Cl1138 is characterized by a significantly higher fraction of
active galaxies and by a less populated (although still relatively tight)
colour-magnitude relation. Among its spectroscopic members, no galaxy shows
signs of recent truncation of star-formation history. At $z\sim 0.5$, cl1232
and cl1138 are the only two clusters for which unbiased and complete
spectroscopic samples are available down to $I(r=1\arcsec)\sim23$. The measured
numbers of k+a galaxies are low (nominally zero for cl1138), and the available
data exclude a large population of post-starburst galaxies at faint magnitudes.

Post-starburst galaxies represent a class of galaxies in the process of
transition from blue actively star-forming population to red quiescent
galaxies. Therefore, they are of significant interest in constraining the
physical processes driving the progressive build-up of the colour magnitude
relation observed in our previous work, and later confirmed by independent
studies using different data-sets. We recall that strong Balmer absorption
lines appear when the star formation in a galaxy is switched off
rapidly. Spectra with strong Balmer lines ($>5$~\AA), require the truncation to
be preceded by a short-lived starburst. K+a spectra with moderate Balmer line
intensities, like most of those measured in our study, might represent
post-starburst galaxies in a later evolutionary stage, but can also be produced
by truncation of `normal' star formation activity. Therefore, the data
presented here allow us to test one specific scenario in which new additions to
the red-sequence are galaxies that have had their star-formation histories
`truncated' by the hostile cluster environment in their recent past.

Through simple calculations, we have shown that the observed low numbers of k+a
galaxies do not appear sufficient to account for the measured increase in the
faint red-sequence galaxy population.  This conclusion remains valid even when
accounting for the infall of new galaxies onto the cluster, and adopting the
very conservative assumption that these all migrate towards the faint-end of
the red-sequence. These arguments are even stronger in the case of cl1138,
which does not contain any spectroscopic member with a k+a spectrum. Although
they contribute at some level (and for some clusters) to the observed evolution
of the faint red-sequence, post-starburst galaxies do not represent the main
channel for the production of new faint red-sequence galaxies at intermediate
redshift. If this population does originate from the blue galaxies observed in
distant clusters, as we speculated in our previous work, the `quenching' of the
star formation probably occurs on longer time-scales (i.e. it is not a
truncation) or, alternatively, the signature of a post-starburst phase might be
too weak to be detected in our spectroscopic sample with the adopted
classification.

In a recent study, \citet{Wild_etal_2008} used a principal component analysis
of the spectra from the VIMOS VLT Deep Survey (VVDS) to identify galaxies with
strong Balmer absorption lines, and estimated that galaxies that have passed
through a strong post-starburst phase account for $\sim 40$ per cent of the
growth in the red-sequence at $z<1$. A direct comparison with our results is,
however, difficult: we focus on a restricted redshift interval and on the
cluster environment, while the results of Wild et al. refer to the general
`field' red-sequence, and provide an estimate of the contribution of
post-starburst galaxies to the red-sequence growth for $z<1$. In addition, it
is difficult to compare our simple classification of k+a galaxies based on the
equivalent widths of the [OII] and H$\zeta$ lines with the more sophisticated
method adopted in \citet{Wild_etal_2008}. We note also that `pre-processing'
happening in other environments (e.g. groups of galaxies infalling onto the
cluster) could help to reconcile our results with those of Wild and
collaborators.

Our results are based on only two clusters with large differences in their
spectral population, and we know that the fraction of post-starburst galaxies
can show large cluster-to-cluster variations
\citep{Poggianti_etal_2009}. Because of bad weather, our spectroscopic
programme could not be completed, and the quality of our spectra is too low to
perform a detailed spectroscopic analysis. Higher quality spectra can be
obtained at this redshift and for the magnitude range considered in this
study. With these data, it might be possible to detect signs of recent star
formation activity in galaxies that we have classified as `passive'. It is
clear that a larger sample of clusters with deeper spectroscopy would be needed
to quantify the importance of the post-starburst phase in building-up the
observed colour-magnitude relation in galaxy clusters.

\section*{Acknowledgements}
Based on observations collected at the European Southern Observatory, Chile, as
part of programme 073.A-0216. The Dark Cosmology Centre is funded by the Danish
National Research Foundation. The Millennium Simulation databases used in this
paper and the web application providing online access to them were constructed
as part of the activities of the German Astrophysical Virtual Observatory. GDL
thanks Nina Novak for precious advice on data reduction and Vivienne Wild for
useful discussions, acknowledges the hospitality of the Kavli Institute for
Theoretical Physics of Santa Barbara, where part of this paper was written, and
financial support from the European Research Council under the European
Community's Seventh Framework Programme (FP7/2007-2013)/ERC grant agreement
n. 202781. IRS acknowledges support from the STFC.

\bsp

\label{lastpage}

\bibliographystyle{mn2e}
\bibliography{dwarfs_gdl}

\end{document}